\documentclass [a4paper,12pt]{article}
\usepackage {graphicx,amsmath}
\begin{document}
\vspace{-15mm}
\begin{center}
 {\Large\bf  The assumption in Bell's inequalities and 
              entanglement problem } \\[2mm]
            Milo\v{s} V. Lokaj\'{\i}\v{c}ek    
\\
{\it Institute of Physics of the AS CR, v.v.i., 18221 Prague 8, Czech Republic } \\
\end{center}

{\bf Abstract } 

 Bell derived the given inequalities on the basis of one rather forceful assumption that was supposed to hold in the hidden variable theory. However, this assumption has been so strong that it has corresponded only to the classical physics; any dependence on spin orientations of individual objects (i.e., on quantum characteristics) having been practically excluded. And consequently, the hidden-variable theory has not been refused by corresponding experimental data. 
 On the other side, Einstein's criticism of Copenhagen quantum mechanics based on ontological approach refusing nonlocality has been entitled as a much higher limit value corresponds to Bell's probability combination in such a case.  And there is not any reason to believe in direct interaction between matter objects at greater distances as shown recently also by K. Hess on the basis of another approach.
 \\   [2mm]

   The Copenhagen quantum mechanics proposed by Bohr \cite{bohr1} is still regarded as the only theory of the microscopic physical processes. It differs very much from the classical physics applied generally to the macroscopic physical world. Bohr's theory was criticized by Einstein et al. \cite{ein} who demonstrated on the basis of a Gedankenexperiment that the mutual interaction between two different matter objects at very great distances was involved in the Copenhagen theory; the phenomenon being now known as the entanglement. However, the given criticism was refused by Bohr \cite{bohr} who stated that the given phenomenon might exist in the microscopic world.  
  
   Later two quantum alternatives were discussed: Bohr's Copenhagen quantum mechanics and hidden-variable theory corresponding to Einstein's requirement. And in 1964 Bell \cite{bell} derived some inequalities that should have held for a slightly modified EPR Gedankenexperiment. It has been assumed that the coincidence detection probabilities of two particles with opposite spins running in opposite directions have being established for differently oriented polarizers in individual detectors:
          \[   \left\|<--|^{\beta}---o---|^{\alpha}-->\right\| \; . \] 

To obtain the given inequalities Bell introduced the assumption that individual detection probabilities were rather strongly mutually correlated. And it was commonly assumed that the given inequalities were valid  in the hidden-variable theory (without any deeper analysis), which seemed to be supported by the fact that the same inequalities were derived in several different ways \cite{clau}. 
 
   And when the experiments (see Ref. \cite{asp}) performed in 1982 showed that Bell's inequalities were violated the hidden-variable theory has been refused and the Copenhagen quantum mechanics has been taken as the only theory of microscopic world even if the passage from this theory to the classical physics of the macroscopic world has not been understood until now. 
   
   The given questions seem to be newly solved at the present; see, e.g., \cite{rosin}-\cite{khren}. At the recent international conference "Frontiers of Quantum and Mesoscopic Thermodynamics 11" held in Prague in July 2011 K. Hess has argued then that there are some hidden assumptions in proofs of Bell's inequalities  and that the claims that Bell's theorem proves influences at a distance are incorrect. In the following we shall analyze the given problem with the help of the method of Bell's operator (see \cite{hill}) to show that the known assumption of Bell is fully responsible for the given conclusion.
   
   In the experiments performed in 1982 the coincidence transmission probabilities of two equally polarized photons through two differently oriented polarizers were established; and the validity of Bell's inequalities 
\begin{equation}
             B = a_1b_1+a_2b_1+a_1b_2-a_2b_2 \leq 2      \label{ineq}
\end{equation}
was tested. The quantities $\,a_j, b_k\,$ have represented transmission probabilities of individual photons passing through two polarizers at different angles between photon spin and polarizer axis. 

Bell derived these inequalities on the basis of a rather strong assumption. He assumed that the value of $B$ for any given combination of four individual probabilities remained unchanged even if one pair of these probabilities (in one polarizer) was interchanged. It was required practically for coincidence probabilities not to depend on spin orientations. However, the physical connotation of this assumption has not been analyzed in the past.    
The inequality (\ref{ineq}) was derived, of course, also in other ways (see Ref. \cite{clau}), but practically the same assumption was always involved in all approaches \cite{lk98}. And it has been commonly assumed that the given inequalities have been valid in the hidden-variable theory without any deeper analysis having been done. 

  Bell's inequalities (\ref{ineq}) were experimentally violated (the upper limit of $2\sqrt{2}$ having been approximately reached), which was interpreted as decisive refusal of the hidden-variable theory and as the victory of the Copenhagen quantum mechanics. 
  However, it has been shown with the help of the already mentioned method of Bell's operator (see \cite{hill}) that several different limits may be obtained for the given combinations of probabilities according to chosen conditions (see, e.g., \cite{revz}). However, an attempt to interpret these different limits physically was undertaken only recently (see \cite{conc2}).  
 
In the mentioned approach the individual probabilities $a_j$ and $b_k$ are to be substituted by operators representing individual measurement acts and acting in two different subspaces (corresponding to individual polarizers) of the whole Hilbert space 
\begin{equation} 
  {\mathcal H}\;=\; {\mathcal H}_a \otimes {\mathcal H}_b \,.   \label{tens}
\end{equation}
It holds for the expectation values of these operators (see \cite{hill})
\begin{equation}
        0\;\leq\; |\langle a_j\rangle|, \;|\langle b_k\rangle|\; \leq \;1\, .  
\end{equation}

Eq.~(\ref{ineq}) represents then the definition of Bell's operator $B$, the expectation values of which may be derived from: $<\!B\!>\;=\;\sqrt{<\!B^+B\!>}$.
 And it may be immediately seen that it must hold $\,<\!B^*B\!>\;\le 16\,$ and consequently \mbox{$\,<\!B\!>\;\,\le\,4\,$}. However, this limit cannot be practically reached; in fact three lower limit may exist according to chosen commutation relations between $\,a_j\,$ and $\,b_j\,$:
\begin{equation}     
     <\!B\!> \;\;\; \leq \;\;\;  2\sqrt{3}, \;\; 2\sqrt{2}\;\;\mathrm{or}\;\;2. 
\end{equation}

The first limit (and actually the highest one derived in \cite{revz} for the first time) corresponds to the case when none of the probability operators commute
\begin{equation}  
      [a_j,b_k]\neq 0\;, \;\;\;[a_1,a_2]\neq 0, \;\;[b_1,b_2]\neq 0\, ,    
\end{equation}
i.e., when the interaction at distance might exist as the operators from different Hilbert subspaces do not commute mutually.

The second limit corresponds to the hidden-variable theory, when only the operators belonging to the same subspaces do not commute (no interaction at distance or no entanglement exists), i.e., if
\begin{equation}  
    [a_j,b_k]= 0\;\;\;\;\mathrm {and}\;\;\;[a_1,a_2]\neq 0,\;[b_1,b_2]\neq 0\,.   
\end{equation}

And finally, the third limit corresponds to the case, when all operators $a_j$ and $b_k$ commute mutually, i.e., if
\begin{equation}  
      [a_j,b_k]= 0 \;,\;\;\;\;\;[a_1,a_2]\;=\;[b_1,b_2]\;= 0  \,,
\end{equation}
which corresponds to the Gedankenexperiment proposed originally by Einstein et al. It is a purely classical (i.e., deterministic) case where any quantum (and probabilistic) characteristics  are not involved. 

The limit values in the last two cases are different even if their actual physical contents differ only rather little. In both these cases the individual objects exhibit locality (no entanglement exists); the only difference between them consists in that in the former case some quantum characteristics (here, e.g., spin orientation) may be involved. The latter case describes the strict classical picture. 
The nonlocality (or the entanglement) exists then in the first case only when the probability operators corresponding to different objects do not commute.  

In the past, as already mentioned, the given limits have not been practically interpreted as to  the  physical content (see, e.g., Ref. \cite{revz}). However, it is evident that only the classical alternative may be regarded as excluded by the results of the EPR experiments. As to the hidden-variable theory it should be denoted in principle as preferred since in the corresponding experiments the value $\;2\sqrt{2}\;$ has been practically filled up but not overpassed. 

 There is not any reason (as argued also by Hess) to speak about the existence of any interaction at a distance (or entanglement) between microscopic objects. However, any hidden assumption in Bell's inequalities need not be considered.  
The requirement of the invariance of any coincidence probability at the interchange of two individual transmission probabilities in one polarizer (as required by Bell) has been fully responsible for the given limit. 

And the physics of the microscopic world may start from the same ontological basis on which the classical physics has been based. The difference consists only in the existence of quantized states (energy as well as angular momentum values of bound systems)
as it follows from corresponding Schr\"{o}dinger equations. Otherwise,  the hidden-variable theory may lead to the same results as the classical physics if the mathematical results are correspondingly physically interpreted \cite{adv}; i.e., only states characterized by individual Hamiltonian eigenfunctions are interpreted as "pure" physical states and any superposition of theirs as a "mixed" state. However, at the difference to Bohr's interpretation the individual eigenfunctions may represent divers instantaneous pure states according to time-dependent Schr\"{o}dinger solutions they belong to (representing the basis vectors of different mutually orthogonal Hilbert subspaces);
all necessary details may be found in Ref. \cite{lkm}.
   
{\footnotesize

\end{document}